\documentclass[final,3p,times,twocolumn]{elsarticle}
\usepackage[dvipsnames]{xcolor}

\usepackage{amssymb}

\journal{Current Opinion in Systems Biology}

\begin{document}

\begin{frontmatter}



\title{Sustainable Diversity of Phage-Bacteria Systems}


\author[inst1]{Namiko Mitarai}
\affiliation[inst1]{organization={The Niels Bohr Institute, University of Copenhagen},
            addressline={Blegdamsvej 17}, 
            city={Copenhagen},
            postcode={2100 \O}, 
            country={Denmark}}
\author[inst1]{Anastasios Marantos}
\author[inst1]{Kim Sneppen}


\begin{abstract}
Bacteriophages are central to microbial ecosystems for balancing bacterial populations and promoting evolution by applying strong selection pressure. Here we review some of the known aspects that modulate phage-bacteria interaction in a way that naturally promotes their coexistence. We focus on the modulations that arise from structural, physical, or physiological constraints. We argue they should play roles in many phage-bacteria systems providing sustainable diversity. 
\end{abstract}



\begin{keyword}
virulent phage \sep kill the winner \sep colony \sep spatial refuge \sep dormancy
\PACS 0000 \sep 1111
\MSC 0000 \sep 1111
\end{keyword}

\end{frontmatter}


\section{Introduction}
\label{sec:intro}
\begin{figure}[h]
\includegraphics[width=0.45\textwidth]{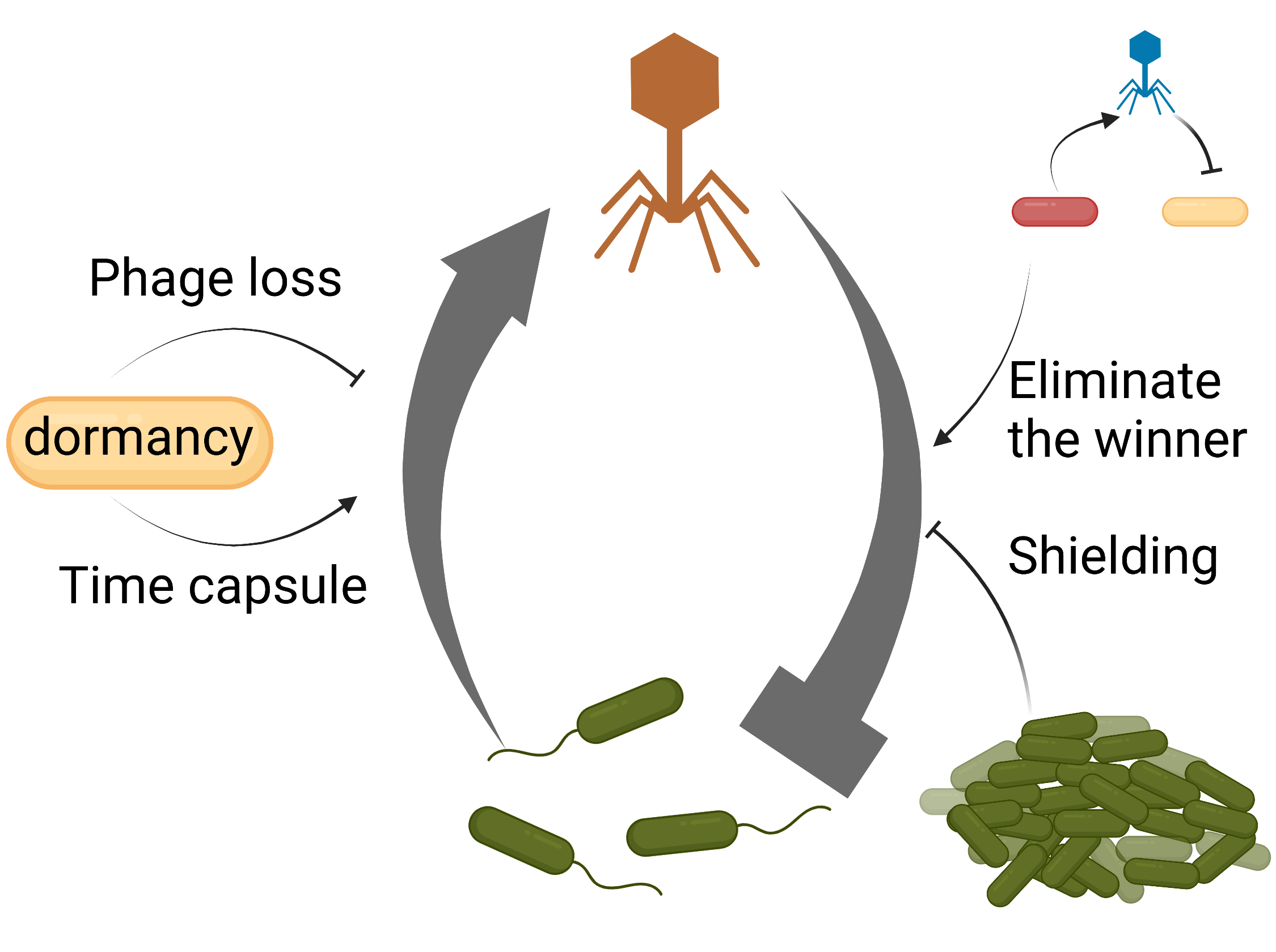}
\caption{Modulation of phage-bacteria interactions for sustainable diversity. Created with Biorender.com. }
\label{concepts}
\end{figure}

Bacteriophages play a critical role in balancing and sometimes reshaping microbial ecosystems, supporting the diversity of bacterial strains \cite{thingstad2000elements,Thingstad2014,haerter2014phage,marantos2022kill} and in facilitating the transmission of genes between different bacterial strains \cite{dixit2015recombinant} and species \cite{koonin2013virocentric}. 
Genetic engineering and gene-editing in modern biotechnology are by-products of
the fitness gain of bacteria when they protect themselves from phages by restriction-modification enzymes \cite{eriksen2022emergence,tesson2023synergy} or CRISPR systems \cite{jinek2012programmable,ishino2018history,rostol2019ph,tesson2023synergy}. Future explorations in phage resistance mechanisms will likely unravel other aspects \cite{scholl2005escherichia,song2020primary,chaudhry2020mucoidy,laughlin2022architecture,millman2022expanded} of the aeon-long war between these dominating life forms \cite{whitman1998prokaryotes,breitbart2005here} on our planet. Here we will review parts of this vast area of research that modulate these interactions to make the system more sustainable in the long term (Fig.~\ref{concepts}), focusing on 1) phages' ability to reshuffle optimal bacterial growth, 2) the effect of spatial refuges against unlimited phage predation, and  
3) implications of the phage lysis dependence on the physiological state of hosts.

\section{Eliminating the winner: ongoing reshuffling of bacterial fitness}\label{sec:eliminate}
\begin{figure*}[h]
\includegraphics[width=\textwidth]{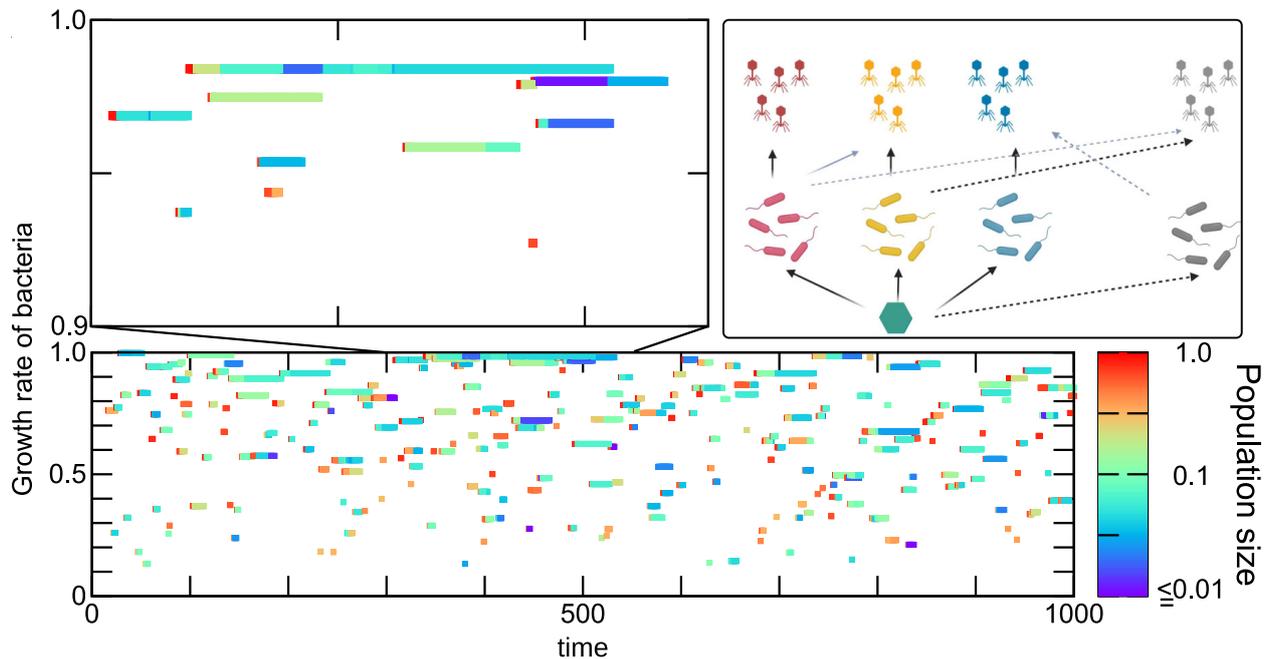}
\caption{From Kill the winner to Eliminate the winner. The plot shows the time evolution of the coexisting bacterial species' growth rate (vertical axis) and population size (colour bar). The lower panel shows a long time course, and the upper panel shows the magnification of the indicated duration. The box figure shows a schematic description of the model (Created with Biorender.com). There is one nutrient source for bacteria feeding all the bacterial species, and phages can predate on one or more bacterial species with variable strength (solid allows). New phage or bacterial species are added to the system one by one randomly after the system reaches the steady state. The predation link between the newly added and existing species (dotted arrows) is drawn randomly. The details of the model are given in \cite{marantos2022kill} as model R.  
}
\label{TasosFig}
\end{figure*}

The resource competition and resulting exclusion make it hard for bacterial species to coexist when they require similar nutrients for growth \cite{gause1934struggle, hardin1960competitive}.
This challenges the observation of diverse microbial communities. Several mechanisms have been proposed, including the cross-feeding of metabolites among different species \cite{seth2014nutrient,goldford2018emergent,goyal2018diversity}, a variation of affinity to slightly different nutrient sources \cite{wang2021complementary}, and spatial in-homogeneity \cite{tilman1994competition,yanni2019drivers}.

This situation changes dramatically once we consider bacteriophages \cite{Thingstad2000,marantos2022kill,haerter2014phage,weitz2013phage,jover2015multiple,xue2017coevolution}. 
Virulent phage specialized in predating on bacterial species keeps the bacterial species population down, and the growth rate of the bacteria is only reflected in the phage population size \cite{campbell1961}, making the nutrient available for more slowly-growing bacterial competitors that otherwise cannot coexist. This phenomenon is called ``Kill the Winner" \cite{Thingstad2000}, and it predicts that slow and fast-growing bacterial species stably co-exist. Notably, however, the slower-growing bacteria are exposed to elimination by invading faster-growing bacteria, provided these invaders sometimes have time without phage predators. 
An open system with an occasional invasion of new bacteria and phage species tends to evolve towards a state with faster-growing bacterial species intermittently \cite{haerter2014phage, haerter2018theory,marantos2022kill}.

However, if individual phage species can prey on several bacteria species, the long-term prediction deviates from this scenario.
Such cross-interactions make the apparent competition \cite{holt1977predation} possible, where the slowly growing bacteria may out-compete the faster grower \cite{marantos2022kill} by being less susceptible to the common phage, 
even when the bacterial species compete for the nutrient source. This is illustrated in Fig.~\ref{TasosFig}, where we follow the
development of an open microbial ecosystem as a function of time counted as the number of bacterial or phage species that enter the system \cite{marantos2022kill}. The model assumes {\it one} common food source for all species. It is simulated in the 
quasi-steady state approximation where new species are only added when the
temporal dynamics of previous species are settled to a steady state. The interaction between the existing species and the newly added species is assigned randomly, with the possibility of cross-interactions (Fig.~\ref{TasosFig}).
Figure~\ref{TasosFig} illustrates the growth rate for each co-existing bacteria, with a colour 
that marks the bacterial population in units of total carrying capacity. 

Figure~\ref{TasosFig} shows that 1) even bacterial hosts with very high growth 
rates can be eliminated, 2) newly introduced bacteria often settle at relatively high 
populations soon after the introduction and subsequently tend to lose population with additions of other competing strains or phages, and 3) High populations of any particular strain are highly transient, while states with about 10\% or lower population density stay longer 
without big changes.

While this model analyzed a completely open system where newly added species properties are chosen independently of existing species, new species may appear due to mutation in a closed system. Hence, their properties are correlated with existing species \cite{lenski1985constraints}. Such a coevolution model gives a nested-modular structure to the resulting interaction network \cite{beckett2013coevolutionary}. A recent study following a coevolving wild population has indicated a local adaptation, i.e., phages are best at infecting the co-occurring hosts \cite{piel2022phage}.
Interestingly, it has been shown that occasional immigration of phages from outside of a subsystem accelerates the local adaptation by introducing more genetic variation \cite{morgan2005effect}. For a quantitative understanding of phage-bacteria ecosystem data \cite{moebus1981bacteriophage,kauffman2022resolving}, a complete approach would be combining local co-evolution with invasion from outside of the system.

\section{Space mediated defence: micro-colony and self-organized spatial refuges against phages}
The above consideration of co-existence suffers from the classical limitations of
ecosystem stability for well-mixed systems described by differential equations
\cite{may1972will}. However, in real ecosystems, forming spatial refuges 
is crucial \cite{tilman1994competition}. For example, coral reefs support a huge diversity of life by providing hiding places against larger predators. The diverse possibility of how a life form can provide hiding for some life forms from their competitors makes formal modelling difficult. One such toy model was to consider the non-transient interactions between lichen species
on a rock surface \cite{mathiesen2011ecosystems}, predicting the existence of 
stable states with high diversity when each species directly interfere only a fraction of other species and if the fraction is below a threshold. 
\begin{figure}[h]
\includegraphics[width=0.45\textwidth]{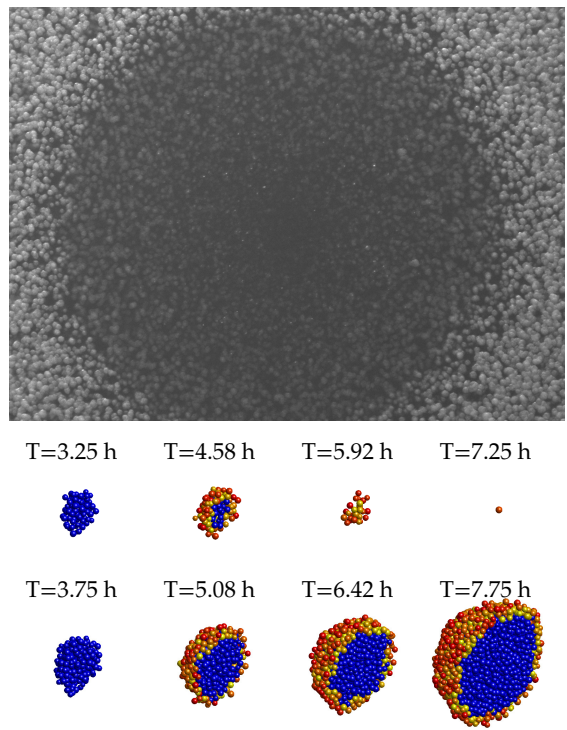}
\caption{Role of microcolonies and space. The top image visualizes plaque formed by CI$^-$ mutant of the $\lambda$ phage, which acts as a virulent phage when infecting a sensitive {\it Escherichia coli} cell (Image from \cite{mitarai2016population}). The white does are bacterial colonies. The killing zone by phage (plaque) appears as a darker region. The radius of the plaque is about one millimetre. The bottom panel shows a simulation of a microcolony attached by a phage from outside (Cross sections are shown, image from \cite{eriksen2018growing}). Blue spheres are uninfected bacterial cells, and yellow to red spheres are the infected cells in the latency period. Phage particles are not visualized. When the first phage attack is early (i.e., the colony is small, top panel), all the cells can eventually be killed. In contrast, when the colony size upon the first phage attack is large enough, the colony can keep growing despite the continuous killing by the phage on the surface. }
\label{microcolony}
\end{figure}

Even a sub-millimetre scale of spatial heterogeneity is sufficient for a microbial system to provide a spatial refuge. Here we consider phages interacting with bacteria on a colony boundary \cite{heilmann2012coexistence, bull2018phage}. In particular, we study bacterial colonies' survival when exposed to a virus \cite{mitarai2016population,eriksen2018growing}. 
One sees a plaque in Fig~\ref{microcolony} formed on a bacterial lawn initially infected by a single virulent phage \cite{mitarai2016population}. Here, a small number of phages were mixed with bacteria in a soft agar, cast on a hard agar with nutrients, and incubated. Subsequently, most single bacteria grew into small colonies since the agar was too viscous for bacteria to be motile. Some
bacteria were attacked by the phage or by its progeny as phages grow and spread by diffusion 
between the growing micro-colonies. Noticeably, increasing remnants of phage-attacked
colonies are observed as one approaches the periphery of the plaque from the inside. This indicates that
larger colonies upon phage attack provide some resistance against phage proliferation, and thus that
bacteria inside a colony are partly protected. 

In \cite{eriksen2018growing}, we infected colonies of various sizes 
with a virulent mutant of the temperate bacteriophage P1. We found 
that colonies larger than $\sim 10^5$ cells at the first phage attack
survived and grew under phage predation, while smaller colonies could not. 
The possible mechanism is illustrated in agent-based simulations in Fig.~\ref{microcolony}.
The phages are modelled as point particles that diffuse in space and infect a bacterial cell upon encountering it. As a result, cells on the colony surface are infected at high multiplicity \cite{taylor2017emergence}, and phages cannot diffuse long inside a colony before they are adsorbed. The colony becomes surrounded by infected and dying bacteria while its non-infected inside grows. When the growth of the bacteria inside the 3-dimensional colony exceeds the death on the 2-dimensional surface layer, the colony can survive. Thus, a tipping point for colony survival primarily depends on phage adsorption and phage latency time relative to the bacterial growth rate \cite{eriksen2018growing}.

The collective protection from phage attack in a colony is empowered further when the bacteria produce extracellular structures that constitute a large part of the biofilm \cite{hansen2019big}. For example, 
the collective protection by expression of curli polymer that can trap T7 phage enables matured enough {\it E. coli} biofilm to keep growing under phage attack \cite{vidakovic2018dynamic,bond2021matrix}; more complex collective protection has been reported recently in a multispecies biofilm \cite{winans2022multispecies}. A simulation study confirmed that the mobility of phage is the key when attacking a biofilm \cite{simmons2018phage}.

From a larger perspective, a medium with many colonies will naturally be able to sustain phage attacks and be much more robust than the homogeneously mixed system. While the dense colony appears as a large target for phage to encounter by diffusion, the overall adsorption rate will be reduced compared to all the cells being dispersed uniformly in the media \cite{abedon2012spatial, eriksen2020phage}. The more the phage encounter is delayed, the larger the microcolony becomes, providing better collective protection. When effectively simulating the shielding of phage attack at the colony surface and the phage diffusion between colonies, robust survival of bacteria and phage have been predicted when bacteria grow as microcolony \cite{eriksen2020sustainability}. 

\section{How to deal with an inactive host: dependence on the host's physiological condition}
Since the production of the phage depends on the infected host's ability to replicate genetic materials (DNA or RNA) and produce proteins (though some phages bring in genes for machinery), the phage growth inevitably depends on the host's physiological states. It has been demonstrated that phage production decreases as the bacterial growth rate slows down \cite{ricciuti1972host,haywood1974lysis,hadas1997bacteriophage,middelboe2000bacterial, you2002effects,abedon2009bacteriophage}. 
A mathematical analysis has demonstrated that such physiology dependence can support a new mode of coexistence between bacteria and virulent phage \cite{weitz2008alternative}.
It is necessary to consider the phage production dependence on the bacterial growth rate to reproduce the plaque morphology and final size in a mathematical model \cite{mitarai2016population}. However, it is worth noting that exceptions such as T7 can keep spreading in a stationary phase loan at a limited speed \cite{yin1992replication}.  Since bacterial growth is limited in various natural environments, it is crucial to investigate the physiology dependence to fully understand the phage-bacteria ecosystem and evolution. 

Interestingly, the host physiology dependence is not limited to the phage yield after infection. Recently, we found that a phage preferentially adsorbs to metabolically active bacteria \cite{brown2022protection}. 
More specifically, the wild type $\lambda$-phage adsorption dropped significantly when the host cell was metabolically inactive, possibly by detecting the hyperdiffusion of the receptor protein in growing \textit{E. coli} \cite{oddershede2002motion} that ceases on energy depletion \cite{winther2009effect}.
This phenomenon requires a particular design of the wild type $\lambda$-phage's tail protein that allows the virus to bind reversibly, which is lost in a \ensuremath{\lambda} host range mutant \cite{schwartz1975reversible}. The preference to infect metabolically more active cells may increase overall phage growth in an environment where hosts with different physiological states are accessible. 

Should a phage always avoid injecting its genetic material into a metabolically inactive host? Of course, infecting a dead host cell will be a pure waste for a phage. However, infecting a dormant but alive host \cite{lennon2011microbial,kolter2022bacteria} may have a long-term benefit. 
An interesting example is a phage which infects bacteria such as {\it Bacillus subtilis} that can form an endospore under certain stressed conditions. Endospores are metabolically inert but can survive under extreme stresses and germinate to regrow when the condition is right again after an immense period \cite{errington2003regulation}.  Many phages that infect spore-forming bacteria can form a "virospore", i.e., an endospore where the phage genome is also encapsulated \cite{sonenshein2006bacteriophages,schwartz2022phage}. When a virospore germinates, the phage can undergo the lytic cycle to produce phage progeny. 
Another example is a phage infecting a dormant persister {\it E. coli} cell, where 
a lytic phage stayed silent while the host was in dormancy, but as soon as the host resumed its growth, phage replication also resumed to complete the lytic cycle \cite{pearl2008nongenetic}.  It has been proposed \cite{stewart1984population,maslov2015well,gandon2016temperate,correa2021revisiting} that survival in a very low host density environment is a possible selection pressure for phage to be temperate, i.e., a phage can enter the lysogeny \cite{howard2017lysogeny} where it stays dormant, and its genetic material can be replicated with the host replication.
Even though the "pseudolysogeny"-like \cite{los2012pseudolysogeny} behaviour by infecting dormant cells does not provide immediate growth of the phage population, it may serve as a time capsule for a virulent phage to survive a harsh environment for a phage particle.

\section{Outlook}
\label{sec:outlook}
We have presented a few aspects of phage-bacteria interactions that make the coexistence of many species more sustainable (Fig.~\ref{concepts}). The "Eliminate the winner" due to possible interaction network structure opens for slow growers to remain in the ecosystem. A dense bacterial colony provides "Shielding" for bacteria deep inside the colony by cells on the surface, absorbing most of the phages attacking from outside. The phage proliferation dependence on host physiology may reduce the impact of phage attack when infected dormant cells do not produce progeny ("Phage loss"). At the same time, if a phage can resume proliferation when an infected dormant cell resuscitates, infecting a dormant cell may work as a "Time capsule" for the phage where the cell and phage stay dormant together and preserved through a crisis that is hard for a phage to survive alone. 

These modulations are not specific to certain molecular mechanisms and therefore expected to be effective in phage-bacteria ecosystems in general.  In addition, players in these systems exhibit many other strategies 
that makes the system even more sustainable. 
We have briefly mentioned the lysis-lysogeny choice in temperate phage \cite{stewart1984population,maslov2015well,gandon2016temperate,correa2021revisiting}; such temperate phage may themselves carry genes that prevent invasion of other phage types \cite{zheng2018type} or genes that allow using other phage genomes to produce offspring (pirate phages \cite{lindqvist1993mechanisms,christie2012pirates,mitarai2020pirate}). 
Many phage defence mechanisms and anti-phage defence mechanisms obviously contribute to sustaining the phage-bacteria coexistence and diversity arising from coevolution \cite{rostol2019ph,eriksen2022emergence,millman2022expanded, tesson2023synergy}. The fascinating diversity of molecular mechanisms challenges the theoretical microbial ecology to provide a unified picture to help us navigate different systems' individuality.

\section*{Conflict of interest statement}
Nothing declared.
\section*{Acknowledgement}
This work was supported by Novo Nordisk Foundation (NNF21OC0068775).
 \bibliographystyle{elsarticle-num} 







\end{document}